\documentstyle[aps,prl,twocolumn]{revtex}
\begin{document}
\draft

\title{Overscreened Single Channel Kondo Problem}

\author{Anirvan M. Sengupta and Yong Baek Kim}
\address{Bell Laboratories, Murray Hill, NJ 07974}

\date{May 16, 1996}

\maketitle

\begin{abstract}

We consider the single channel Kondo problem with
the Kondo coupling between a spin $S$ impurity
and conduction electrons with spin $j$.
These problems arise as multicritical points
in the parameter spaces of two- and higher-level
tunneling systems, and some impurity models of heavy 
fermion compounds. In contrast to the previous 
Bethe-anstaz conjectures, it turns out that the 
dynamics of the spin sector is the same
as that of a spin $S$ impurity coupled to $k(j)$ channels 
of spin $1/2$ electrons with $k(j) = 2j(j+1)(2j+1)/3$.
As a result, for $2S < k(j)$, the system shows non-Fermi liquid
behavior with the same exponents for the thermodynamic quantities
as those of $k(j)$ channel Kondo problem.
However, both the finite-size spectrum and the operator content
are different due to the presence of the other sectors 
and can be obtained by conformal field theory techniques. 

\end{abstract}

In the ordinary single channel Kondo problem,
the conduction electrons with spin $1/2$ interact with a  
spin $1/2$ impurity\cite{kondo}. 
This model has been regarded as a canonical model for the
physics of dilute magnetic impurities in metals\cite{kondo}. 
The physics of the model can be summarized as the complete 
screening of the impurity spin by the conduction electrons 
in the low energy limit, which leads to the Fermi-liquid 
behavior of the physical quantities\cite{kondo}.
Nozi\`eres and Blandin\cite{nozieres} pointed out that 
the above picture of the single channel Kondo model should 
be changed when the number of channels, $k$, for the 
electrons with spin $1/2$ is larger than twice the impurity 
spin $S$. This overscreened multichannel Kondo model 
has been the subject of intensive 
research\cite{andrei,wiegmann,al,ek,gogolin}
due to its non-Fermi liquid behavior.     

The purpose of this paper is to study the single 
channel Kondo model in which the conduction electrons 
with spin $j$ interact with the spin $S$ impurity.
These models arise as multicritical points\cite{vladar}
in two-level systems proposed by Zawadowski and 
Vladar\cite{zawadowski}. Recently, Moustakas and Fisher 
consider the more general models, for example, 
systems with three impurity positions\cite{fisher}. 
Our models are realized as some special fixed points, 
with enhanced symmetry, in their parameter space.
According to Kim and Cox, $j=3/2$ case arises 
in the context of the theory of heavy fermions with
$Ce^{3+}$ impurities in the cubic environment\cite{cox}.

In the classic paper of Wiegmann and Tsvelik\cite{wiegmann}
on multichannel Kondo problems, it was conjectured that the 
$k$-channel Kondo problem is equivalent to the single channel 
Kondo problem with conduction electron spin $k/2$\cite{sacramento}. 
We will see that this is not correct. We believe that the 
discrepancy is due to the fact that the Bethe-ansatz approach
does not take care of the bulk degrees of freedom correctly. 
So far, the only reliable method for studying these models 
have been finding fixed points in the limit of a large number 
of flavors\cite{zarand}. Our results are not only consistent 
with what was known, but also provide exact scaling dimensions 
around these fixed points.
 
We show that the dynamics of the spin sector in this problem
is the same as that of a spin $S$ impurity coupled to 
$k(j)$ channels of spin $1/2$ electrons with $k(j) = 2j(j+1)(2j+1)/3$.
As a result, for $2S < k(j)$, the system shows non-Fermi liquid
behavior with the same exponents for the thermodynamic quantities
as those of $k(j)$ channel Kondo problem.
However, it turns out that the finite size spectrum of 
the system is different from that of the $k(j)$ channel Kondo
problem due to the presence of the remaining degrees of 
freedom and can be obtained by conformal field theory techniques.   

In the Kondo problem, it is sufficient to consider only 
the s-wave scattering or the radial motion of the conduction 
electrons.
This one-dimensional problem in half-space ($x \ge 0$ and $x$ 
is the radial coordinate) can be represented in terms
of left-going (chiral) fermions $\psi_{\alpha} (x)$ defined for 
$-\infty < x < \infty$.
The Hamiltonian of the conduction electrons with spin $j$ interacting
with the spin $S$ impurity,
$H = H_0 + H_1$, can be written as
\begin{eqnarray}
H_0 &=& i v_F \sum_{\alpha} \int^{\infty}_{-\infty} dx \
\psi^{\dagger}_{\alpha} (x) \partial_x \psi_{\alpha} (x) \cr
H_1 &=& \sum_{a = x,y,z} {\cal J} S^a J^a (0) \ ,  
\label{ham}
\end{eqnarray}
where $\alpha = j, j-1, ... , -(j-1), -j$, $S^a$ are the
impurity spin $S$ operators, and $J^a (x)$ are the conduction 
electron spin currents or densities:
\begin{equation}
J^a(x) = \sum_{\alpha \beta} \psi^{\dagger}_{\alpha} (x) 
t^a_{\alpha \beta} \psi_{\beta} (x) \ ,
\label{spcurrent}
\end{equation} 
where $t^a$s are spin $j$ representations of $SU(2)$ generators. 

First we show that the dynamics of the spin sector is the same
as that of a spin $S$ impurity coupled to $k(j)$ 
channels of spin $1/2$ electrons with 
\begin{equation}
k(j) = 2j(j+1)(2j+1)/3 \ .
\label{level}
\end{equation}
This relies on the fact that, in the Kondo perturbation theory,
the only quantities which enter are the multipoint correlators
of the electron spin densities at the impurity site.
The long time behavior of these correlators is specified by
the level $k$ of the current algebra satisfied by the spin
currents in the equivalent one dimensional problem.
The level $k$ can be determined from the two-point function of
the currents:
\begin{equation}
\langle J^a(x) J^b(y) \rangle = {k/2 \over (x-y)^2} \delta^{ab} \ .
\label{twopoint}
\end{equation}
Here $k$ is given by $k(j) = 2 {\rm Tr} (t^a t^a) = {2 \over 3} \sum_a 
{\rm Tr} (t^a t^a) = 2j(j+1)(2j+1)/3$.
Therefore, the spin sector is decribed by $SU_{k(j)} (2)$ 
Wess-Zumino-Witten (WZW) model\cite{kz}.

The interaction of the electrons with the impurity spin 
can be incorporated by the changes in the boundary condition of the 
$SU_{k(j)} (2)$ WZW model\cite{al}.
In the case  of the $k$ channel Kondo model for $k \ge 2S$, the change
in the boundary condition corresponds to the fusion with the spin $S$
operator of $SU_k (2)$\cite{al}.      
Since the Hamiltonian in the spin sector of the present problem 
looks exactly the same as that of the $k(j)$ channel Kondo problem, 
we expect that at the conformal fixed point the boundary condition 
is provided by the fusion with the same operator.
Therefore, for $S < j(j+1)(2j+1)/3$, we have overscreening of the
impurity spin and non-Fermi liquid behavior even in the 
single channel problem.

In order to consider the full Hilbert space, we have to consider
other degrees of freedom as well as the spin sector.
First, we consider the $U(1)$ charge sector with the corresponding 
charge current $J(x) = \sum_{\alpha} \psi^{\dagger}_{\alpha} (x) 
\psi_{\alpha} (x)$.
One can write down $SU_1(2j+1)$ current algebra for the other 
degrees of freedom by considering the currents of the form
$J^A = \sum_{\alpha \beta} \psi^{\dagger}_{\alpha} 
T^A_{\alpha \beta} \psi_{\beta}$, where $T^A$s are general Hermitian 
traless $(2j+1) \times (2j+1)$ matrices.
Only three of these currents are needed to describe the spin
degrees of freedom and form $SU_{k(j)} (2)$ current algebra.
Thus the rest of the degrees of freedom other than spin and charge 
can be formally described by the coset model $SU_1(2j+1)/SU_{k(j)}(2)$.
Note that the central charge of this coset model is 
$c_{\rm coset} = 2j - 3k(j)/[k(j)+2]$.

To be more specific, let us consider $j=1$ and $j=3/2$.
For $j=1$, $k(1) = 4$ (which was discussed in \cite{gogolin}) 
and the coset model is trivial in the sense 
that there is no degree of freedom ($c_{\rm coset} = 0$). 
For $j=3/2$, $k(3/2) = 10$ and the coset model corresponds to 
single Majorana fermion or equivalently the critical 
Ising model ($c_{\rm coset} = 1/2$).
This can be shown more explicitly by using Abelian 
bosonization as follows.

For $j = 3/2$, the four fermion operators can be bosonized as\cite{emery}
\begin{equation}
\psi_{\alpha} (x) = {1 \over \sqrt{2 \pi a}} e^{-i\Phi_{\alpha}(x)} \ ,
\end{equation}
where $\alpha = 3/2, 1/2, -1/2, -3/2$ and $a$ is a short-distance cutoff
of the order of the lattice constant.
Let us consider
\begin{eqnarray}
\Phi_c &=& {1 \over 2}(\Phi_{3/2} + \Phi_{1/2} + \Phi_{-1/2} + \Phi_{-3/2}) \cr
\Phi_s &=& {1 \over 2 \sqrt{5}}
(3 \Phi_{3/2} + \Phi_{1/2} - \Phi_{-1/2} - 3 \Phi_{-3/2}) \cr
\Phi_t &=& {1 \over 2 \sqrt{5}}
(\Phi_{3/2} - 3 \Phi_{1/2} + 3 \Phi_{-1/2} - \Phi_{-3/2}) \cr
\Phi_b &=& {1 \over 2}(\Phi_{3/2} - \Phi_{1/2} - \Phi_{-1/2} + \Phi_{-3/2}) \ .
\end{eqnarray}
The spin currents can be written as
\begin{eqnarray}
J^z &=& \sqrt{5} \ \partial_{x} \Phi_s \cr
J^+ &=& {1 \over \pi a} (\sqrt{3} \ {\rm cos} \ \Phi_b \
e^{i(\Phi_s + 2 \Phi_t)/\sqrt{5}} + 
e^{i(\Phi_s - 3 \Phi_t)/\sqrt{5}}) \ ,
\end{eqnarray}
where $J^{\pm} = J^x \pm i J^y$.
Note that the spin currents involve only $\Phi_s$, $\Phi_t$, and
a Majorana fermion $\chi_1 = {\rm cos} \ \Phi_b$.
The boson $\Phi_c$ and the Majorana fermion 
$\chi_2 = {\rm sin} \ \Phi_b$ decouple from the spin sector.
The first three degrees of freedom provide the central charge
$c = 5/2$ as expected for $SU_{10} (2)$.
$\Phi_c$ corresponds to the charge sector and the remaining
Majorana fermion $\chi_2$ is the coset degree of freedom
refered previously.

In conformal field theory, there is one-to-one correspondence
between operators and states. Each of these operators can be
considered as a product of operators coming from three sectors
described earlier. 
This is called the conformal embedding\cite{al}.
The above mentioned fusion will affect only the operators in the spin 
sector through the formal operator product expansions.
The new operators found after fusion can be mapped to a well 
defined set of states of the same conformal field theory.
These states form the Hilbert space of the interacting problem.

As the first step, one constructs the states in the finite-size 
free electron Hilbert space and writes down the corresponding 
operators in terms of operators coming from the three sectors.
Let us consider the free electrons moving on a finite-size 
one-dimensional lattice, which can be regarded as the 
regularization of the continuum model.
In the absence of any potential, this model has the particle-hole
symmetry. Let us be at the half-filling and preserve the symmetry.
In this case, for even number, $2m$, of sites, the chemical potential, 
$\mu$, lies exactly in the middle of $m$th and $(m+1)$th levels and all 
other levels are symmetically placed around $\mu$.
On the other hand, if we had odd number, $2m+1$, of sites, the chemical
potential would lie exactly on the $(m+1)$th level and the remaining
levels would be placed symmerically around it.

Let us consider the continuum model given by $H_0$ with $x$ 
restricted to the interval $-L/2 < x < L/2$.
If the boundary condition is 
$\psi(L/2)=e^{2i\delta}\psi(-L/2)$, all the energy levels with respect to
the chemical potential are quantized as
$E_n = v_F k_n = {2 \pi v_F \over L}(n + {\delta \over \pi})$.
For the particle-hole symmetric case, $\delta$ is either $0$ or $\pi/2$.
Note that $\delta = 0$ corresponds to the case of odd number of sites
in the lattice problem while $\delta = \pi/2$, to the case of even number
of sites. 
We will call the Hilbert spaces corresponding to
these two cases as the periodic sector (P) and the 
anti-periodic sector (AP).
In principle, the energy spectra of P and AP sectors can be 
compared with those of the odd and the even iterations in numerical
renormalization group calculation.
 
As an example of overscreened Kondo problem, we calculate 
the finite-size spectrum for the case of $j = 3/2$ and $S = 1/2$. 
Let us start with AP sector. 
After the conformal embedding, some of the low lying 
states for $j=3/2$ and their quantum numbers are given in the Table I.
Here $j_{\rm tot}$ refers to the total spin of each state. 
In case that this state is a descendent in the spin current algebra, we
indicate it by putting prime ($'$). 
In the second column, we indicate the conformal primary from the Ising
sector (Majorana fermion)\cite{al}.
$q$ refers to the total charge and the prime again indicates 
the descendent in the charge current algebra.
The last two columns provide the excitation energy (in units of 
$2 \pi v_F / L$) with respect to the ground state energy
$\Delta_0$ and the corresponding degeneracy of the state.

The spin $j_{\rm tot}$ primaries of $SU_{k(3/2)} (2)$ have dimension 
$j_{\rm tot}(j_{\rm tot}+1) / [k(3/2) + 2] = 
j_{\rm tot}(j_{\rm tot}+1) / 12$.
In the Ising sector, the dimensions of $\epsilon$ and $\sigma$
are $1/2$ and $1/16$ respectively.
The charge sector primaries with charge $q$ have dimension
$q^2/8$. 
In each sector, the decendents can be obtained by repeatedly applying 
Fourier modes of the spin current $J^a (x)$, the charge current $J (x)$, 
and the stress tensor $T (x)$ of the Ising sector respectively.
Applying the $n$th Fourier mode increases the dimension by $n$.
The total energy (in units of $2\pi v_F / L$) of the state is 
given by the sum of the dimensions from the three sectors.

Now we introduce the interaction with the impurity spin $S$. 
At the new fixed point, after the interaction is included, 
the states in the Hilbert space of the interacting problem 
can be also written in terms of the constituent degrees of freedom
from the spin, the charge, and the coset sectors.
In order to get the Hilbert space of the interaction problem, 
we use the general algebraic procedure suggested by 
Ref.~\cite{al,cardy}, 
which is to fuse the operator in the spin sector with the spin 
$S$ operator of $SU_{k(j)}(2)$. 
If the operator from the spin sector has spin $j'$, 
then the fusion generates operators of
$|j'-S|, |j'-S|+1, ... , {\rm min}(j'+S, k(j)-j'-S)$.
Following this procedure in the case of $j = 3/2$ and $S=1/2$, we
get the Table II from the Table I and it represents the spectrum 
of the interacting problem.
For convenience, we list only states with excitation energy
below $2 \pi v_F / L$ (this corresponds to 
$\Delta_{\rm tot} - \Delta_0 < 1$).
Note that for none of the tables we try to list all 
the primary operators in the spectrum, which can be done
in a straightforward manner. 

For example, the second row of Table I has a $j_{\rm tot} = 3/2$
primary in the spin sector.
Fusing with the spin $1/2$ primary of $SU_{10} (2)$ gives rise
to spin $1$ and spin $2$ primary operators.
These provide the second and the third row of the Table II,
where only the spin quantum numbers are changed.

For the case of P sector, following the same procedure, 
we obtain the Table III and the Table IV which 
correspond to the noninteracting and the interacting spectrum 
respectively. 
Here we also list only the first few levels in the
interacting spectrum.   

All the singular behaviors in the thermodynamic quantities are
due to the leading
irrelevant boundary operator, which can be added 
to the Hamiltonian.  
Here this leading irrelevant operator
is exactly the same as that in the $k(j)$ channel 
Kondo problem, {\it i.e.}, the first descendent of
the spin $j_{\rm tot} = 1$ operator\cite{al}.
As a result, it has the dimension 
$1 + \Delta = 1 + {2 \over k(j)+2}$, which is $7/6$ for $j=3/2$.
Here $\Delta \equiv {2 \over k(j) + 2}$ is the dimension of the spin
$j_{\rm tot} = 1$ operator.
Accordingly the susceptibility and the specific heat coefficient 
diverge as $T^{2 \Delta -1} = T^{-2/3}$ as 
in the $10$ channel Kondo model\cite{al}. 

Let us investigate various perturbations around the fixed point.
In particular, we are interested in $SU(2)$ symmetry breaking
interactions between the impurity spin and the electrons, 
which is of the form 
\begin{equation}
H_{\rm pert} = \sum_{a=x,y,z} S^a \psi^{\dagger}_{\alpha} 
\Lambda^a_{\alpha \beta} \psi_{\beta} \ .
\label{pert}
\end{equation} 
Here $\Lambda^a_{\alpha \beta}$ are also 4x4 traceless Hermitian
matrices which are orthogonal to spin $j$ matrices in the sense 
that ${\rm Tr}(\Lambda^a t^b) = 0$.

In order to classify possible relevant boundary operators
(which correspond to adding various interaction terms   
at the impurity site),  
we perform double fusion\cite{al,cardy} and 
look at the operator content.
We need to consider only operators which are bosonic, charge 
conserving, and have dimension less than one. 
There are three such operators, which are 
${\cal O}^a$, ${\cal O}^{(ab)}$,
and ${\cal O}^a \times \epsilon$. 
Here ${\cal O}^a$ is a spin 1 operator which has dimension 1/6
and couples to an external magnetic field.
${\cal O}^{(ab)}$ is a traceless symmetric matrix and corresponds
to a spin 2 operator which has dimension 1/2.
The last operator ${\cal O}^a \times \epsilon$ 
involves the energy operator $\epsilon$ of the
Ising sector so that it has the total dimension $2/3$.  
In the absence of the external magnetic field, but in the 
presence of $SU(2)$ breaking terms which are of the form 
in Eq.~\ref{pert}, the second operator can be generated.
Since this is a relevant operator, this fixed point will be 
unstable against such a perturbation.

However, as we consider the case of $S > 1/2$ and $j > 1$, 
we find that there is an $SU(2)$-invariant charge-conserving   
relevant operator. Which discrete symmetries can rule out this 
operator is not clear to us. This operator,
if present, can take us to a new $SU(2)$-invariant fixed point. 

In summary, we study the single channel Kondo 
problem in which the conduction electron with spin $j$ interact
with a spin $S$ impurity.
It is shown that the dynamics of the spin sector is the same
as that of $k(j)=2j(j+1)(2j+1)/3$ channel Kondo problem 
in which conduction electrons with spin $1/2$ of $k(j)$ channels 
interact with a spin $S$ impurity. 
Thus, for $2S < k(j)$, the system shows non-Fermi liquid 
behavior. As a result, the exponents of the specific heat 
coefficient and the impurity susceptibility should be 
the same as those of the $k(j)$ channel Kondo problem.
However, it is also pointed out that the finite-size
spectrum should be different due to the presence of 
the remaining degrees of freedom. 
As an example, we show the low lying states in finite-size 
spectrum of the system in the case of $j = 3/2$ and $S = 1/2$ for
the periodic and the anti-periodic sectors
We also analyze the possible relevant perturbations around
the fixed point.

{\it Acknowledgements} We thank Natan Andrei for discussions about
the Bethe-ansatz conjecture. We also thank D. Cox for helpful comments.  

{\it Note Added}\ : After this work was completed, we were informed
that T.-S. Kim, L. Oliveira, and D. L. Cox studied the model with the
conduction electron spin $3/2$ using the numerical renormalization 
group.\cite{cox2} We also received a preprint by Fabrizio and 
Zarand\cite{fabrizio}, which arrived at the same conclusion as ours.

\begin{table} 
\caption{The energy spectrum for non-interacting electrons with 
spin $j = 3/2$ in AP sector.} 
\begin{tabular}{lllll} 
$j_{\rm tot}$ & Ising & $q$ & $\Delta_{\rm tot} - \Delta_0$ & Deg. 
\\ \hline
\tableline 
0 & 1 & 0 & 0 & 1 \\ \hline
3/2 & $\sigma$ & $\pm 1$ & 1/2 & 8 \\ \hline
2 & 1 & $\pm 2$ & 1 & 10 \\
0 & $\epsilon$ & $\pm 2$ & 1 & 2 \\ \hline
2 & $\epsilon$ & 0 & 1 & 5 \\
3 & 1 & 0 & 1 & 7 \\ \hline
$1'$ & 1 & 0 & 1 & 3 \\ 
0 & 1 & $0'$ & 1 & 1 \\
\vdots & \vdots & \vdots & \vdots & \vdots \\ 
\end{tabular} 
\end{table}

\begin{table} 
\caption{The energy spectrum for interacting system with $j = 3/2$
in AP sector.} 
\begin{tabular}{lllll} 
$j_{\rm tot}$ & Ising & $q$ & $\Delta_{\rm tot} - \Delta_0$ & Deg. 
\\ \hline
\tableline 
1/2 & 1 & 0 & 0 & 2 \\ \hline
1 & $\sigma$ & $\pm 1$ & 7/24 & 6 \\ \hline
2 & $\sigma$ & $\pm 1$ & 5/8 & 10 \\ \hline
5/2 & 1 & 0 & 2/3 & 6 \\ \hline
3/2 & 1 & $\pm 2$ & 3/4 & 8 \\
3/2 & $\epsilon$ & 0 & 3/4 & 4 \\ 
\vdots & \vdots & \vdots & \vdots & \vdots \\
\end{tabular} 
\end{table}

\begin{table} 
\caption{The energy spectrum for non-interacting electrons with 
spin $j = 3/2$ in P sector.} 
\begin{tabular}{lllll} 
$j_{\rm tot}$ & Ising & $q$ & $\Delta_{\rm tot} - \Delta_0$ & Deg. 
\\ \hline
\tableline 
0 & 1 & $\pm 2$ & 0 & 2 \\ \hline
3/2 & $\sigma$ & $\pm 1$ & 0 & 8 \\ \hline
0 & $\epsilon$ & 0 & 0 & 1 \\
2 & 1 & 0 & 0 & 5 \\
\vdots & \vdots & \vdots & \vdots & \vdots \\
\end{tabular} 
\end{table}

\begin{table} 
\caption{The energy spectrum for interacting system with $j = 3/2$
in P sector.} 
\begin{tabular}{lllll} 
$j_{\rm tot}$ & Ising & $q$ & $\Delta_{\rm tot} - \Delta_0$ & Deg. 
\\ \hline
\tableline 
3/2 & 1 & 0 & 0 & 4 \\ \hline
1 & $\sigma$ & $\pm 1$ & 1/24 & 6 \\ \hline
1/2 & 1 & $\pm 2$ & 1/4 & 4 \\ \hline
1/2 & $\epsilon$ & 0 & 1/4 & 2 \\ \hline
2 & $\sigma$ & $\pm 1$ & 3/8 & 10 \\
\vdots & \vdots & \vdots & \vdots & \vdots \\
\end{tabular} 
\end{table}

\end{document}